# Approximate analytical method and its use for calculation of phase velocities of acoustic plane waves in crystals for example $LiNbO_3$

*A.A. Golubeva*

*Saratov State Technical University,*

*Polytechniceskay 77, Saratov, 410054, Russia*

*E-mail: annamoroz@mail.ru*

## Abstract

*By means of the offered analytical method the determinant relation for a phase velocities of elastic waves for an arbitrary propagation directions in a piezoelectric crystal are received. The phase velocities of three normal elastic waves for the crystal of $LiNbO_3$ are calculated. Results of this calculation for each of waves are presented graphically in the form of the cards allowing easily to define phase velocities in any given direction in crystal.*

In a crystal velocities of distribution of elastic waves and their polarization depend on a distribution direction about the crystallographic axes. Generally at any given direction of normal line three waves with various phase velocities and mutually orthogonal vectors of polarization can extend. The wave, which vector of displacement makes the least angle with a direction of distribution, is called as quasilongitudinal. Two other waves are define as lateral excitation. The definition of phase velocities of these three waves, and also their different parametres, is the basal problem of acoustics. The main purpose of this report are calculations of phase velocities of elastic waves in $LiNbO_3$ on the basis of the offered new analytical coordinate method. The offered method assumes generally occurrence piezoelectricity and is not limited by any class of symmetry, it differs favorably a generality and simplicity the received results from the existing method which basis on the equations of Christoffel and mathematical methods of their decision [1]. The results received for $LiNbO_3$ which is known for a long time and studied enough, are in good conformity with the data received by a traditional method, for example in [2]. Full analysis of phase velocities of elastic waves in some crystals can appear rather useful to research of possibility of use of this crystal for production of acousto-electronic and acousto-optical devices.

It is possible to enter into consideration a new system of coordinates $x', y', z',$ where axis $z'$ is coincide with a direction of distribution of plane elastic waves. Rotation of coordinate system can be defined by two angles $\varphi$ and $\theta$. By means of the equations of movement, the material equations and Maxwell`s equations which have been written down in tensorial form, in this system of coordinates, the system of the equations about different component of a vector of mechanical displacement has been received

$$\rho \frac{\partial^2 u'_i}{\partial t^2} = \overline{c'}_{i33m} \frac{\partial^2 u'_m}{(\partial x'_3)^2}; \tag{1}$$

where: $u'$ - is the component of the vector of displacement in the coordinate system connected with a direction of distribution of wave; $\rho$ - density of the excited medium; $\overline{c'}_{i33m}$ - tensors of coefficient of elasticity in displaced crystallographic system of coordinates, and these coefficients "are toughened" by piezoeffect; $x'_3$ - the direction of distribution of plane elastic waves.

Elements of tensors $\|c\|, \|e\|, \|\varepsilon\|$, defining elastic, piezoelectric and dielectric properties of the medium have been converted from original system of rectangular coordinates in new one, connected with any given direction in a crystal. Further such new system of coordinates $x''y''z''$, in which elastic waves are normalised, i.e. the mechanical displacement corresponding to each of three independent waves are directed on axes of this new system is found. This new system of coordinates can be defined by three independent angles $\delta, \chi, \psi$ , which are required. After the second transformation of coordinates, the system of the equations (1) will be converted to three wave equations independent from each other, each of that defines this or that elastic wave:

$$\rho \frac{\partial^2 u_i''}{\partial t^2} = \bar{c}''_{i33m} \frac{\partial^2 u_m''}{(\partial x_3')^2}; \tag{2}$$

$$\bar{c}''_{i33m} = \bar{c}'_{k33n} \alpha_{ik} \alpha_{mn}; \tag{3}$$

where: $|\alpha|$ - matrix of transformation to the second system of coordinates.

Then the criterion of normalisation of required waves will look like:

$$\bar{c}''_{i33m} = 0, если\ i \neq m,\ \bar{c}''_{i33i} = c_{i3}^{эф}, \tag{4}$$

where: $c_{i3}^{эф}$ - three effective coefficients of elasticity, which define phase velocities of three normal elastic waves according to expression:

$$v_s = \sqrt{c_{i3}^{эф} / \rho}\ . \tag{5}$$

Thus, angles $\delta, \chi, \psi$ are defined at first, and then effective coefficient of elasticity and phase velocities of elastic waves are found. For the first wave disclosing of expression (2) taking into account known angles $\delta, \chi, \psi$ gives:

$$c_{13}^{эф} = [c_{55} \cos^2 \eta + c_{44} \sin^2 \eta + c_{45} \sin 2\eta] - [c_{33} - c_{55} \cos^2 \chi - c_{44} \sin^2 \chi - c_{45} \sin 2\chi] \times$$

$$\times \cos^2 \psi \sin^2 \delta + [(c_{55} - c_{44}) \sin 2\chi - 2c_{45} \cos 2\chi] \sin 2\psi \sin^2 \delta / 2. \tag{6}$$

Last two summands in this expression are sizes of the second order of smallness about a small angle $\delta$. They should be considered, if the split-hair accuracy in effective coefficients definition is necessary, and, accordingly very much for phase velocities of elastic waves. If them does not consider, this coefficient is defined by the first summand in expression, and this expression taking into account formulas for angles can be led to a kind:

$$c_{13}^{эф} = 1/2(c_{55} + c_{44}) + c; \qquad (7)$$

where:

$$c = \sqrt{\frac{1}{4}(c_{55} - c_{44})^2 + c_{45}^2}, \qquad (8)$$

For the second wave, operating similarly, it is possible to receive expression:

$$c_{23}^{эф} = [c_{55} \sin^2\eta + c_{44} \cos^2\eta - c_{45} \sin 2\eta] - [c_{33} - c_{55} \cos^2\chi - c_{44} \sin^2\chi - c_{45} \sin 2\chi] \sin^2\psi \sin^2\delta +$$

$$[(c_{55} - c_{44}) \sin 2\chi - 2c_{45} \cos 2\chi] \sin 2\psi \sin^2\delta / 2. \qquad (9)$$

It is necessary to notice that the first wave is a quasi perpendicular wave, mechanical displacement in which are directed on an axis $x''$, the second is also a quasi perpendicular wave mechanical displacement in which are directed on an axis $y''$.

For the third quasi longitudinal wave:

$$c_{33}^{эф} = c_{33} + [c_{33} - c_{55} \cos^2\chi - c_{44} \sin^2\chi - c_{45} \sin 2\chi] \sin^2\delta; \qquad (10)$$

By means of the of the offered method phase velocities of three elastic waves for an arbitrary propagation directions in a crystal of $LiNbO_3$ (the trigonal system, class 3m, density of 4,66*$10^3$ kg/$m^3$, elastic, piezoelectric, dielectric constants were are taken from the directory [3]) are calculated. Results of calculation for each of waves are presented graphically in the form of the cards in a stereographic projection, allowing to define values of velocities in any given direction in a crystal with ease. Values of velocities were described in the form of constant lines, and the distribution direction was given by means of angles $\varphi$ and $\Theta$, $\varphi$ changed from 0 ˚to 90 ˚with increment in 5 ˚, and $\Theta$ changed from 0 ˚to 180 ˚with increment in 5 ˚.

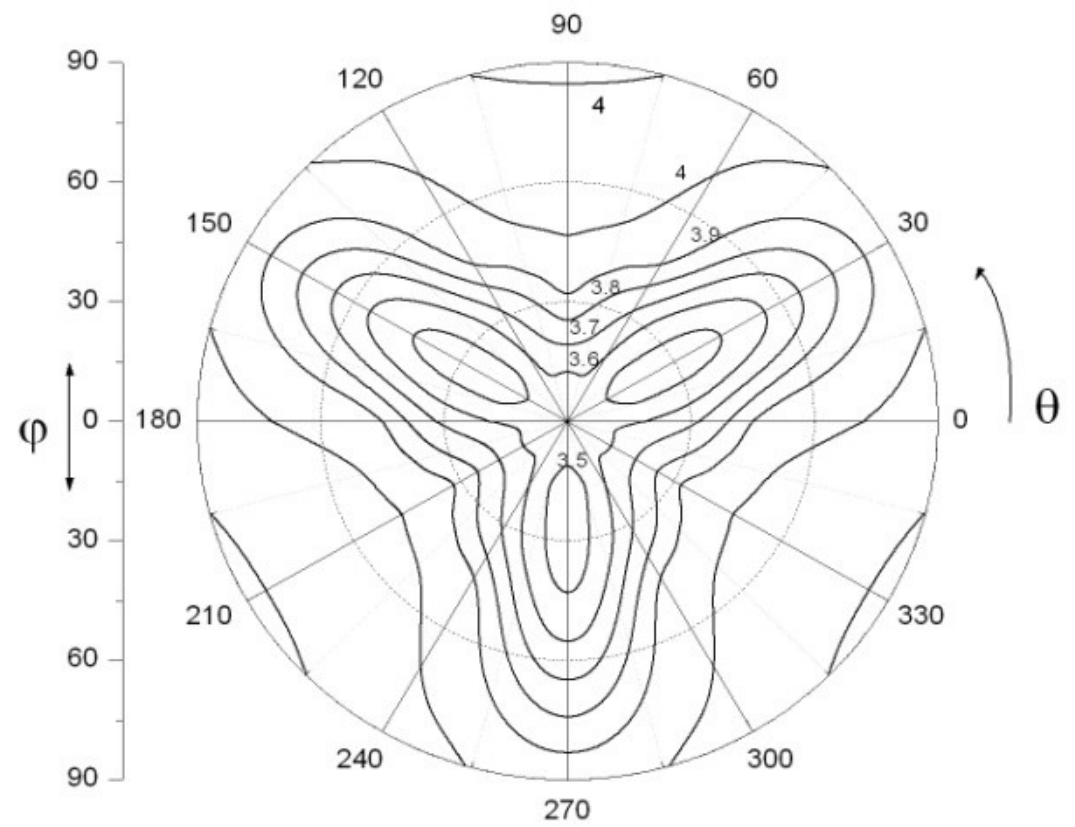

Fig.2 Distribution of phase velocities depending on a direction of propagation for first quasi perpendicular wave, mechanical displacement are directed on an axis $x''$, in a crystal of $LiNbO_3$. Velocities are given in km/s.

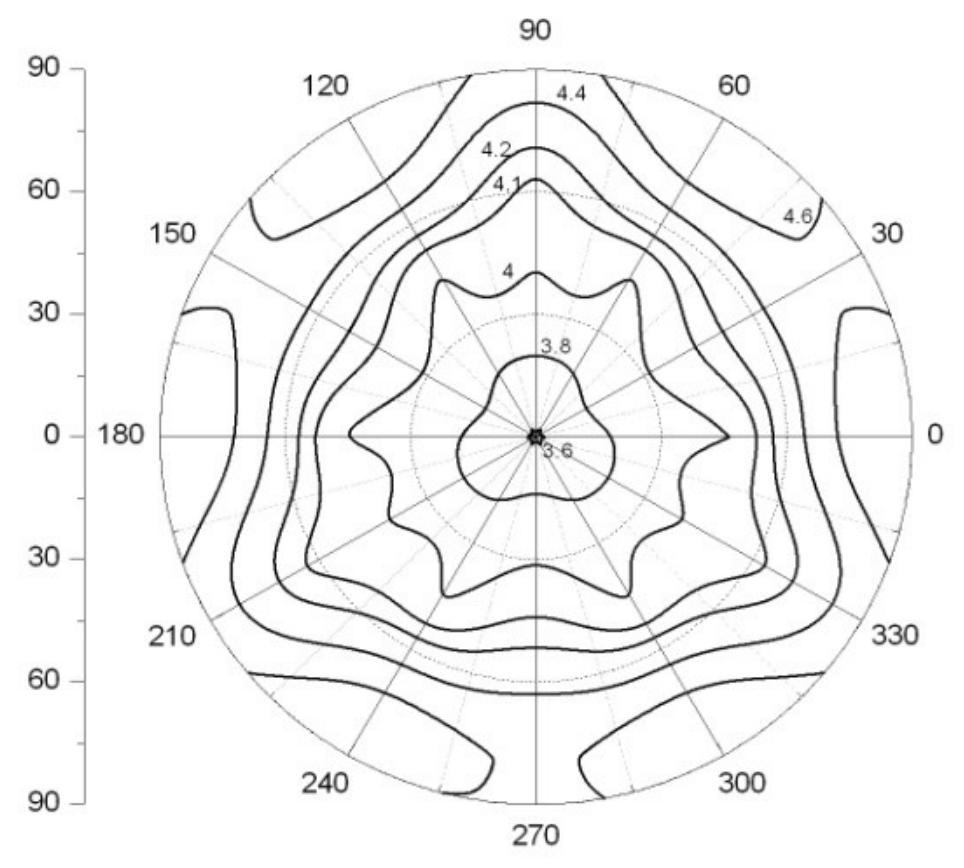

Fig.3 Distribution of phase velocities depending on a direction of propagation for second quasi perpendicular wave, mechanical displacement are directed on an axis $y''$, in a crystal of $LiNbO_3$. Velocities are given in km/s.

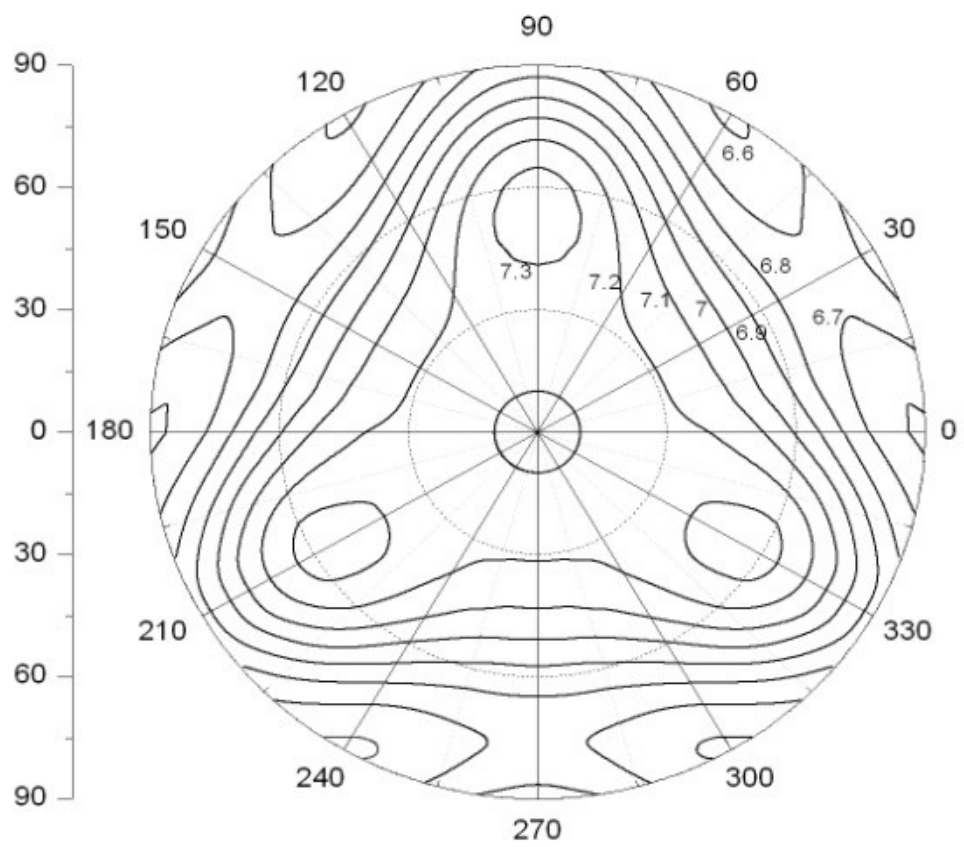

Fig.1 Distribution of phase velocities depending on a direction of propagation for a quasi longitudinal wave in a crystal of $LiNbO_3$ . Velocities are given in km/s.